# Dynamics of Fecal Coliform Bacteria along Canada's Coast

Shuai You[1], Xiaolin Huang[1], Li Xing[3], Mary Lesperance[1], Charles LeBlanc[4], Paul Moccia[5], Vincent Mercier[6], Xiaojian Shao[2], Youlian Pan[2, *], Xuekui Zhang[1, *]

[1] University of Victoria, 3800 Finnerty Road, Victoria BC V8W 2Y2, Canada

[2] Digital Technologies Research Centre, National Research Council Canada, 1200 Montreal Road, Ottawa ON K1A 0R6, Canada

[3] University of Saskatchewan, 105 Administration Place, Saskatoon Saskatchewan S7N 5A2, Canada

[4] Shellfish Water Classification Program - Atlantic Region, Environment and Climate Change Canada, Government of Canada, 443 University Ave., Moncton, NB E1A 3E9, Canada

[5] Shellfish Water Classification Program – Pacific Region, Environment and Climate Change Canada, Government of Canada, 2645 Dollarton Highway, Vancouver BC V7H 1B1, Canada

[6] National Coordination, Environment and Climate Change Canada, Government of Canada, 443 University Ave., Moncton, NB, E1A 3E9, Canada

* Corresponding authors:
  Xuekui Zhang: xuekui@uvic.ca; ORCID: 0000-0003-4728-2343
  Youlian Pan: youlian.pan@nrc-cnrc.gc.ca; ORCID: 0000-0002-0158-0081

Emails of other authors:
  SY: shuaiyou@uvic.ca
  XH: xiaolinhuang@uvic.ca
  LX: lix491@mail.usask.ca
  ML: mlespera@uvic.ca
  CL: Charles.LeBlanc@ec.gc.ca
  PM: Paul.Moccia@ec.gc.ca
  VM: Vincent.Mercier@ec.gc.ca
  XS: Xiaojian.Shao@nrc-cnrc.gc.ca

# Abstract

The vast coastline provides Canada with a flourishing seafood industry including bivalve shellfish production. To sustain a healthy bivalve molluscan shellfish production, the Canadian Shellfish Sanitation Program was established to monitor the health of shellfish harvesting habitats, and fecal coliform bacteria data have been collected at nearly 15,000 marine sample sites across six coastal provinces in Canada since 1979. We applied Functional Principal Component Analysis and subsequent correlation analyses to find annual variation patterns of bacteria levels at sites in each province. The overall magnitude and the seasonality of fecal contamination were modelled by functional principal component one and two, respectively. The amplitude was related to human and warm-blooded animal activities; the seasonality was strongly correlated with river discharge driven by precipitation and snow melt in British Columbia, but such correlation in provinces along the Atlantic coast could not be properly evaluated due to lack of data during winter.

# Introduction

Fecal coliform bacteria are potentially pathogenic microorganisms that originate in the intestines of human beings and warm-blooded animals. They can be transported in fecal excrement to various water environments from sources like sewage effluent as well as wildlife and waterfowl (Ksoll et al., 2007). The contaminated water has been shown to place risk on human health. For example, exposure to fecal contaminated water is associated with an increased risk of diarrhea (Laborde et al., 1993); consumption of water contaminated by fecal coliform bacteria is associated with typhoid fever (Mermin et al., 1999), gastrointestinal illness (Weissman et al., 1976), and hepatitis (Bergeisen et al., 1985). As a result, fecal coliform bacteria levels are regarded as one type of indicator that reflects the fecal contamination level of water (Noble et al. 2003). A high level of fecal coliform bacteria in the water indicates a degradation in the marine environment, and, as such, levels are widely measured to monitor the quality of surrounding water environments across the world, such as in the US (Soueidan et al. 2021), China (Aram et al. 2021), and Korea (Park et al. 2006). The consumption of shellfish harvested from areas contaminated with fecal coliform bacteria increases health hazards, as the contaminants in the seawater are concentrated/biomagnified by the filter feeding bivalve shellfish.

To sustain a healthy bivalve molluscan shellfish industry, the Canadian Shellfish Sanitation Program (CSSP) was established in 1948 to monitor the health of shellfish harvesting areas. As a part of the program, Environment and Climate Change Canada (ECCC) has been monitoring and recording the concentration of fecal coliform bacteria, which are mostly E. coli (Dufour, 1977), along with salinity since 1979. Currently, bacteriological water quality is monitored at an average of 6,000 sites annually, designated for shellfish harvesting across 6 coastal provinces in

Canada: British Columbia (BC), Quebec (QC), New Brunswick (NB), Prince Edward Island (PE), Nova Scotia (NS), and Newfoundland and Labrador (NL). Data from nearly 15,000 unique sites were evaluated in this study. Prior to 1999, there were multi-day surveys conducted in selected locations. After 1999, samples were better distributed over space and time. Over time, ECCC routinely assessed fecal coliform levels at sites and accordingly adjusted shellfish harvesting schedules for closures and re-openings of specific harvesting areas. The assessment typically involved statistical analyses on recent multi-year sample sets of data at particular shellfish harvesting sites, identification of potential sources of point and non-point pollution, and classification of the sites based on their fecal contamination levels. To complement this effort, we conducted a sweeping study across regions using multi-year datasets to provide macro-scale understanding of potential water quality impairment and make further recommendations for coastal management. For better understanding of potential causal factors, river water flow data were retrieved from the Environment Canada Data Explorer developed by the Government of Canada and available. Using this rich data collected over four decades, we aimed to find temporal fluctuation patterns of fecal coliform bacteria levels at sites in each coastal province and their associations with temporal physical climate and oceanographic factors, which would help with more efficient monitoring of coastal marine environmental contamination in Canada.

The temporal bacteria data are of extremely high dimensions due to the large number of distinct time points and locations at which the measurements were taken. Concisely summarizing the association between two sets of high-dimensional temporal data is a challenge. Temporal data often carry substantial information in the correlation between data observed in adjacent time points. Hence, we decided to use lower-dimensional surrogate variables to replace the high-dimensional temporal data. Principal component analysis (PCA) is a popular tool for this type of task, as it finds sets of uncorrelated linear combinations of the input features, i.e., principal components (PCs), that decompose the variance of the features among the samples (Jolliffe and Cadima, 2016). As a result, the originally high-dimensional data of each sample are summarized into a collection of sample-specific values indicating how observations of the sample differ in each of the directions represented by the PCs. However, PCA does not consider the temporal order in the data, which was a key attribute of our data, and PCA requires that the input data is a matrix without any missing data, but our data were collected at sparse and irregular time points that were drawn from a continuum, i.e., a 40-year time period. We reduced the dimension of our data using the non-parametric Functional Principal Component Analysis (FPCA) based on empirical conditional estimation (Wang et al., 2016).

In this paper, we unveil two prominent patterns in the fecal coliform bacteria levels via application of FPCA to data collected in British Columbia. The data characteristics represented by these two patterns and their relationships with specific regional geographical factors are then illustrated through linear regression models and correlation analyses. This workflow was subsequently applied in provinces along the Atlantic coast in Canada, where there was a significant limit on data availability. The structure of the article is as follows. Section 2 provides information regarding the available data, the preprocessing procedure and the methods used. Pattern discovery and characterization, and their correlations with geographical factors are presented in Section 3. Finally, Section 4 provides a discussion, conclusion, and consideration of further research directions.

## 2. Data and Methods

### 2.1 Data

Data consist of measurements of fecal concentrations collected at nearly 15,000 shellfish harvesting sites across 6 coastal provinces in Canada, including British Columbia (BC), Quebec (QC), New Brunswick (NB), Prince Edward Island (PE), Nova Scotia (NS), and Newfoundland and Labrador (NL). At each site, water samples of unit volume were taken for monitoring purposes from 1979 to 2018/2019. The number of the fecal coliform bacteria in each sample (bacteria count per 100 mL of water) was then measured according to the standard method (Braun-Howland et al., 2017). Temperature (ºC) and salinity (‰) were collected along with each bacteria-count measurement. In addition, we obtained two other datasets: 1) precipitation records over the 240-hour period prior to each sampling, and 2) GPS coordinates of the CSSP monitoring sites. In total, the data were in the form of 18 matrices, with 3 for each province.

### 2.2 Data Preprocessing

The amount of available data for each province varied across times and sites. To screen out data that were not suitable for analyzing the variation patterns of fecal coliform contamination levels, the following three exclusion criteria described in You et al (2022) were applied with adjustment to the third criterion. Briefly, we only focused on data collected after 1999, when the sampling was more frequent and regular at each site across the six provinces, and thus sites that did not have any measurements after 1999 were removed. Secondly, we excluded the sites showing no contamination. In this regard, sites with all measurements below 2 fecal coliform bacteria per 100 mL of water sample, which is below the detection limit, were removed.

There were no significant cross-year differences in the bacterial measurements. As the data were collected to serve the CSSP monitoring program, the data density spreading over the 20 years was very low and unreliable to generate meaningful results. To increase the data density, for each site, we pooled the 20 years data onto a scale of one year (365 days) and further reduced the time resolution from 365 days to 52 weeks by taking the average of 7-day periods (You et al., 2022). Data in each weekly bin were summarized using the mean of all counts within the weekly bin. After pooling, we further excluded sites that had consecutively missing measurements over any 4-week period because they would not be reliable to reveal seasonal variation of the bacteria concentrations. Consequently, 847 (20.9%) of the 4062 sites in BC were retained.

Sites in BC tended to have data across the year, whereas the sites in the five provinces on the Atlantic coast did not have data in the winter months. Most measurements were available between late May and mid-November in QC, NB, PE and NS. In NL, the measurements covered an even shorter time period. Therefore, we combined the data in QC, NB, PE and NS into one dataset and focused on the subset that ranged from the $19^{th}$ (mid-May) to the $45^{th}$ week (early mid-November). In NL, we focused on the subset from the $20^{th}$ week to the $38^{th}$ week (early October). After the same screening, 906 (28.2%) of the 3213 sites in QC, 776 (39.0%) of the 1989 sites in NB, 700 (78.2%) of the 895 sites in PE, 1041 (31.0%) of the 3359 sites in NS, and 1006 (78.5%) of the 1282 sites in NL were retained.

A log base 10 transformation was applied to the weekly mean counts before subsequent analyses. The precipitation data corresponding to the sites, as well as the water flow data in recent years of different rivers obtained from Environment Canada Data Explorer [(https://www.canada.ca/en/environment-climate-change/services/water-overview/quantity/monitoring/survey/data-products-services/national-archive-hydat.html)](https://www.canada.ca/en/environment-climate-change/services/water-overview/quantity/monitoring/survey/data-products-services/national-archive-hydat.html) were preprocessed in the same manner, except for the logarithm transformation. Cumulative precipitation over the 5 days prior to each sampling and weekly average water flows were used for downstream correlation analyses.

## 2.3 Dimension Reduction via Functional Principal Component Analysis

We modeled the measurement at the $i$-th site and time $j$ as $X_i(j)$. Here the measurements $X_i(j)$ were treated as realizations drawn from underlying functions and kernel functions are applied during estimation of the FPCA model. According to the application of FPCA on sparse data via the "fdapace" package in R (Gajardo et al., 2021), a moderate amount of missing data on $X_i$ are acceptable. As a result, more available measurement data contribute to more reliable results, while missing measurements on individual samples can be interpolated based on the learned model.

If the data are in the form of a matrix, with each row corresponding to a sample and each column corresponding to a time point, then the functional principal components (FPCs), $\phi$'s, are a set of ordered orthonormal eigen functions, derived one by one, that maximize the variance of $\int_J (X_i(j) - \mu(j))\phi(j), i = 1, \ldots, n$, where $\mu$ is the mean function over the continuum, $n$ is the sample size, and $J$ is the domain of time $j$.

Upon finding the FPCs, an appropriate number, $K$, were selected as the number of FPCs that users considered as an optimal summary of the variance in the samples. In the "fdapace" package, based on the Principal Analysis by Conditional Estimation (PACE) algorithm (Yao et al., 2005), under the assumption that the random measurement errors in the centered data follow a Gaussian distribution with a zero mean and a constant variance, the maximum number of FPCs that can be obtained is $n - 2$. Typically, the FPCs explain decreasing proportions of variance in the centered data, and most of the variance gets explained by several top FPCs. As a result, selecting a small $K$ can result in the loss of significant information, whereas selecting a large K may include in the model patterns related to outliers or noise. Users mostly start from the first FPC and end when a sizeable accumulative percentage of variance has been explained while the remaining FPCs seem to barely convey any useful information. We followed this procedure in this study.

With the FPCs and $K$ determined, the value at the $j$-th time point of the $i$-th sample's function can be well approximated by a linear combination of the $K$ selected FPCs and the corresponding $K$ FPC scores, $\beta_{i,k}, k = 1, \ldots, K$, i.e.,

$$X_i(j) \approx \mu(j) + \sum_{k=1}^{K} \beta_{i,k} \cdot \phi_k(j) \qquad [1]$$

where $\mu$ is the mean function averaged over all functions at every time point, and

$$\beta_{i,k} = \int_J (X_i(j) - \mu(j))\phi_k(j) \qquad [2]$$

is the score assigned to the $i$-th sample's function corresponding to the $k$-th FPC which measures how strong this function differs along the direction of the $k$-th FPC against the mean function. Therefore, most of the information in each sample is summarized into a collection of FPC scores that are the coordinates of the sample in the newly constructed $K$-dimensional FPC space.

## 2.4 Correlation Analysis and Visualization

Linear regression (Seber and Lee, 2012) and correlation analyses (Gogtay and Thatte, 2017) were applied to find the correlations between potential data characteristics and the variation patterns found via FPCA as well as between these patterns and potential environmental factors and other contributors of fecal contamination. We considered two correlations, Pearson correlation, which measures the linear relationship between two variables, and the rank-based Spearman correlation, which instead measures the monotonic relationship. A comparison between the two illustrates the advantage of Spearman correlation when the data have high kurtosis, outliers, or a heavy-tailed distribution (De Winter et al., 2016). Since the data in this study were highly sparse and skewed, and abrupt peaks of bacteria levels were frequently observed without intermediate measurements, Spearman correlation was more appropriate and used throughout this study.

The "ggmap" package in R (Kahle and Wickham, 2013) was used for visualization of the scores in all regions in this study. After the variation patterns and the data characteristics represented by them were found, the FPC scores functioned as indicators of the strengths of those characteristics. Mapping them on the corresponding sites' geographical locations enabled the discovery of potential environmental factors that were correlated with the observed bacteria levels on the sites.

An interactive website, which uses the "shiny" package in R (Change et al., 2015) to visualize both the data and the analytical results, is available at https://opms.uvic.ca and updated periodically.

## 3. Results

### 3.1 Coastal Fecal Coliform Patterns in British Columbia (BC)

The data from the coast of British Columbia (BC) were available across the entire year, so we started to explore the distribution patterns based on this relatively complete dataset.

### 3.1.1 Variance Decomposition of Fecal Coliform Bacteria Levels

Functional Principal Component Analysis (FPCA) was applied to model the variance in the data. The results are presented in Fig. 1. In Fig. 1a, the mean function appeared at its lowest near the end of March while at its highest in November; the top 2 functional principal components (FPCs) explained 95% of the variance from the mean function at the sites, and their variations over time are depicted in Fig. 1b.

The FPC1 explained 74% of the variance. According to the black curve in Fig. 1b, it stayed positive throughout the year, and thus, during approximation of the bacteria levels at a site via Eq. 1, a larger FPC1 score would contribute to a larger amount of the FPC1 value added to the approximation in every week, and we would see an overall increase in the approximated bacteria levels at the corresponding site. Such relationship indicated that the FPC1 reflected the overall amplitude of the bacterial level at a site across a year. Therefore, sites with higher overall amplitudes were assigned larger FPC1 scores.

This finding was supported by a significant correlation between the maximum level of bacteria at a site throughout the year and the percentile of its FPC1 score ($R^2=0.61$, $p<2.2e-16$, Fig. 2a). The maximum values at different sites appeared at different amplitudes and different times across the year, and the separation between the sites with the top and bottom 10% of FPC1 score was obvious (Fig. 2b). These findings suggested that the FPC1 scores indeed reflected the overall amplitude of the bacteria levels at sites in BC.

Apart from the FPC1, the FPC2 accounted for 21% of the variance. It was negative before May and after November, while positive in the middle of a year (Fig. 1b). As a result, the FPC2 scores indicated positive or negative contributions in different seasons of the year as reflected by Eq. 1.

This time-related pattern indicated the seasonality in a year; sites with higher and positive FPC2 scores had their highest bacteria levels during the middle of a year, while sites with lower and negative FPC2 scores had their highest at the beginning or the end of a year.

For a clearer depiction of this seasonality, we looked into the sites with stronger seasonal variations in their bacteria levels among those with the highest overall amplitudes of bacteria levels, i.e., those with the top 10% of the FPC1 scores (the red dots in Fig. 2b). Two groups were extracted based on two extrema (the top and the bottom 10%) of the FPC2 scores. Their approximated bacteria levels according to Eq. 1, represented by red and blue curves in Fig. 3a and Fig. 3b, corresponding to 28 and 20 sites, respectively. These approximations have been validated in more detail (Supplementary Fig. S1). While the seasonal variations of the two groups were drastically different, sites within each group showed synchronized temporal patterns in their approximated bacteria levels across a year. Among sites with the higher FPC2 scores, their highest approximated bacteria levels appeared in summer and fall (Fig. 3a), whereas the sites with lower FPC2 score had low values from spring to early fall (Fig. 3b). Given different ranges (every 10%) of FPC1 scores, such contrast between high or low FPC2 scores was consistent across all amplitudes of bacterial levels represented by the entire spectrum of FPC1 scores (Supplementary Fig. S1). The FPC2 thus justifiably explained seasonal variation patterns in the bacteria level at each site.

### 3.1.2 Amplitudes of bacteria levels around BC coast

With the main variation patterns confirmed, for efficient visualization of the relative overall amplitudes of bacteria levels, sites were grouped into 10 bins of equal size in the order of increasing FPC1 scores. Sequential colors were assigned to these bins, blue for the ones with the lowest FPC1 scores while red for those with the highest (Fig. 4).

According to Fig. 4, many high-amplitude sites appeared in the Strait of Georgia in close proximity to the city of Vancouver, southeast Vancouver Island between Courtenay, Nanaimo, and Victoria, and along the west coast of Vancouver Island, while low-amplitude sites were along the northwest of the mainland side of the strait and in between the small islands from the Texada Island to Desolation Sound (marked as the grey circles in Fig. 4). Such distribution was confirmed by marking the sites with the highest and the lowest 10% of the FPC1 scores, respectively (Supplementary Fig. S2).

The sites around major cities with large populations usually had high FPC1 scores that indicated the overall amplitudes of fecal coliform concentration. Such cases were around Vancouver (662,248 people), Victoria (91,867 people), Nanaimo (99,863 people), and Courtenay (28,420 people), according to the [2021 Census of Population](#) by Statistics Canada. Between Victoria and Nanaimo, some sites appeared low-amplitude. Compared to the high-amplitude sites in this region, locations of the low-amplitude ones were farther away from the coast. This is reasonable since a large proportion of fecal coliform contamination would be expected to originate from inland populated or agricultural areas and travel seaward.

Fig. 4 also shows a number of sites with high amplitudes of fecal coliform contamination on the west coast of Vancouver Island, which are not highly populated, but are subject to some of the highest levels of rainfall in the province. Non-point-source pollution due to human and wild-life activities is usually regarded as the cause of the contamination in such areas. The spring months are replete with wildlife such as birds and marine mammals and upland animals, and certain areas are very popular with boaters, tourists, and outdoor enthusiasts. Contributions from these animal and human activities would bring higher amounts of fecal matter to the area.

### 3.1.3 Seasonality of bacteria levels and leading factors

The seasonality in bacteria levels represented by the FPC2 was also of major interest. Using the same procedure, assigning 10 sequential colors to 10 bins of increasing FPC2 scores resulted in distributions of the FPC2 scores among the sites (Fig. 5). There was a clear contrast in the bacteria levels at the sites in the vicinity of major municipalities, such as the area from Campbell River to Victoria and around Vancouver City, as compared to elsewhere. The sites away from cities usually showed red, where the bacteria levels correlated with the discharge patterns of rivers driven by snow melt, which peaked in summer, from the mountains in BC (Fig. 3c and Fig. 3e). The sites around the major cities appeared to show blue, where the bacteria levels correlated with the regional precipitation pattern (Fig. 3d). The sites located in between the cities on the east coast of Vancouver Island were also mostly blue, which agreed with the discharge patterns of small rivers mostly driven by precipitation that peaked in winter and early spring, followed by a drought in summer, but then increased again in the fall due to increasing rainfall (Fig. 3f).

#### 3.1.3.1 River Discharge

The annual discharge patterns of the Squamish River and the Fraser River highly agreed with the seasonal variation of bacteria levels of the sites with high FPC2 scores (Fig. 3a, c, e). The Squamish River is a proglacial river to the northwest of Vancouver City, driven by the meltwater from the Coast Mountains (Hickin, 1984). It forms one of the many similar paths for the inland water to access the Strait of Georgia, on the southwest coast of the mainland. The surrounding sites at these discharging locations from the north side of Strait of Georgia to the northwest of Vancouver City were dominated by red colors in Fig. 5. According to Fig. 3b, these sites typically had high bacteria levels in the summer and low bacteria levels in winter. The approximated bacteria levels at individual sites in Fig. 3a were highly correlated with Squamish River annual water flow from 1999 to 2018 in Fig. 3c ($\rho \geq 0.74$, $p < 2.2e-16$).

The Fraser River is the longest river in BC. It has a much larger drainage area that flows from some of the warmest parts of BC, including parts of Rocky Mountain, to the Strait of Georgia. Its seasonal flow pattern (Fig. 3e) was similar to the Squamish River's, except that the Fraser River flow appeared to peak a few weeks earlier according to flow data collected at a farther inland location. Approximated bacteria levels at individual sites shown in Fig. 3a were also highly correlated with Fraser River annual water flow patterns from 1999 to 2016 in Fig. 3e ($\rho \geq 0.62$, $p < 1.4e-06$).

*3.1.3.2 Precipitation*

Populated regions around and in between cities in BC are surrounded by blue-ish colors in Fig. 5, indicating low FPC2 scores. Bacteria levels at sites in these regions, resembling those of the sites in Fig. 3b, were high in winter, the season of high precipitation, and low in summer, the dry season in coastal BC.

Moreover, certain river systems are sensitive to or mostly driven by precipitation. On the east side of Vancouver Island, the Courtenay River, Englishman River, Nanaimo River, Chemainus River, and Cowichan River are such examples (Fig. 3f). Sites at the discharging locations of these rivers were mostly covered in blue-ish colors in Fig. 5, which suggested the contribution of precipitation to the high concentrations of fecal matter in the rainy seasons in these areas.

The correlation between the bacteria levels at those sites with the strongest but opposite seasonal variation (sites in Fig. 3b) and the precipitation measured in BC supports this observation. At measuring locations (sectors) corresponding to individual sites, each measurement was the cumulative precipitation over the past 5 days prior to each Fecal Coliform sampling, and the mean of all such measurements within the weekly bins were acquired for analysis. The weekly precipitation levels at sites in Fig. 3b (those with the highest 10% of FPC1 scores and the lowest 10% of FPC2 scores) were presented in Fig. 3d. We obtained the correlation values between the approximated fecal coliform bacteria levels and the approximated precipitation levels (by FPCA, with >95% of variance explained) at those individual sites. The correlations were found all significant ($\rho \geq 0.54$, $p < 5.7e-05$). In Fig. 6, the correlation analysis between approximated levels of fecal coliform bacteria and precipitation was performed at all sites in BC; sites with significantly positive correlations were marked, with the correlations grouped into 10 bins.

Other than the urban areas, some scattered sites at which the bacteria levels were highly correlated with precipitation appeared to the northwest of Vancouver City, the north of Nanaimo, and the east of Courtenay (around Texada Island, marked as the large grey circle in Fig.4). During periods of high precipitation, the bacteria within the upland drainage area may enter river systems and are discharged into the marine where they can become entrained in surface waters.

In summary, the FPCA analyses at the sites in British Columbia indicated that the majority of the variance among their fecal coliform bacteria levels revealed differences in the overall amplitude of fecal contamination in a year as reflected by the FPC1; whereas, the seasonal variation of the contamination level was reflected by the FPC2. The high and low FPC2 scores revealed distinct patterns of seasonal variations. The sites with high FPC2 scores had high amounts of fecal matter in the summer months during seasonal presence of high human/animal activities in the regions or high discharge from rivers which carried fecal material from inland areas. Sites with low FPC2 scores, on the other hand, were affected by regional precipitation or discharge from smaller regional river systems that were sensitive to precipitation. These sites were generally in the urbanized regions or near areas of high agricultural activities.

## 3.2 Fecal Coliform Patterns along Atlantic coast

We applied the same techniques described above to coastal marine environments across Atlantic Canada and analyzed the data in the same manner.

### 3.2.1 Variance decomposition of fecal coliform bacteria in QC, NB, PE, and NS combined, and in NL

Based on data availability, FPCA was applied to the period of week 19 to 45 for the provinces of Quebec (QC), New Brunswick (NB), Prince Edward Island (PE), and Nova Scotia (NS), and to week 20 to 38 in Newfoundland and Labrador (NL) alone (Supplementary Fig. S3).

The FPC1 in the two scenarios explained 89% and 79% of the variance, respectively. The positive values of its variations over all the selected weeks suggested positive contributions in Eq. 1. Therefore, similar to that in BC scenario, every FPC1 score indicated the overall amplitude of the bacteria level at a site in these provinces from 1999 to 2019. The linear relationship between the maximum observed bacteria level at a site and the corresponding FPC1 score (Fig. 7a, c) was found to be significant in both the 4 combined provinces ($R^2=0.68$, $p<2.2e-16$) and in NL alone ($R^2=0.74$, $p<2.2e-16$). Meanwhile, in each scenario, the occurrences of the maximum measurements at sites were quite evenly distributed over the selected weeks (Fig. 7b, d), and there was very little overlap between the maximum measurements collected at sites with the highest or the lowest 10% of FPC1 scores, which justified our reference to an FPC1 as a reflection of the overall amplitude of the fecal contamination at the corresponding site in these 2 scenarios.

The FPC2 explained 7% and 17% of the variance among bacteria levels in the 4 provinces combined and NL alone, respectively. Each FPC2 showed a change from positive to negative values around the 35$^{th}$ week (Supplementary Fig. S3b, d). In the former scenario, accounting for such a small percentage of variance, the approximated bacteria levels at sites with two extrema of FPC2 scores did not present major differences in their variation patterns; in the latter scenario, the differences were larger. Among sites with the 10% highest FPC1 scores, the increasing trend of bacteria levels stopped at those with the highest 10% of FPC2 scores after around the 33$^{rd}$ week, while those with the lowest 10% of FPC2 scores had generally increasing bacteria levels over the range (Supplementary Fig. S4).

The two leading factors, the river discharge and the precipitation that were found correlated with the seasonality in the bacteria levels in BC, were then studied in these provinces along the Atlantic coast (Supplementary Fig. S5).

In QC, NB, PE, and NS combined and in NL alone, the river discharge patterns had similar trends in a year. Some of the stream-gauged rivers were selected in these provinces, and they all peaked around the 20$^{th}$ week, followed by a trough that lasted until around the 40$^{th}$ week. The major rivers, such as the Saint John River in NB and the Churchill River in NL (Figure S5a, c), had increasing flows afterwards, whereas the smaller rivers remained in their trough for the rest of the year.

The precipitation patterns appeared similar in the 2 scenarios (Figure S5b, d). The precipitation was less variable over the selected weeks along Atlantic coast.

The seasonal variations of the bacteria levels and their correlation with corresponding river discharge and precipitation patterns at sites in these provinces over the selected range of time appeared not as prominent as in BC. Given that there were major seasonal data void in the bacteria-count, further analyses on this part would require more complete data. For example, our analysis of fecal coliform data covered the time range from mid or late May to early October in the case of Newfoundland and Labrador, and to mid-November in the case of the other four provinces. This would miss the peak precipitation in spring as it already declined in mid-May. The fall peak of precipitation would be missed as well for the case of Newfoundland and Labrador.

### 3.2.2 Amplitudes of Bacteria levels and Populated Areas in QC, NB, PE, NS, and NL

The FPC1 scores for sites in the 4 provinces combined and NL were both ordered and separated into 10 bins of equal size. The same 10 sequential colors as in Fig. 4 were then assigned to the bins. The distributions of the colors are mapped to the sites' geographical locations in QC (Supplementary Fig. 6), NB (Supplementary Fig. 7), PE (Supplementary Fig. 8), NS (Supplementary Fig. 9), and NL (Supplementary Fig. 10).

The distributions of FPC1 scores in these provinces agreed with that in BC: colors indicating high amplitudes of bacteria levels were usually surrounding populated areas, such as cities, villages, and communities. For example, such areas included Gaspe and Rimouski in QC, Tracadie, Caraquet, and Bathurst in NB, Alberton, Summerside, Souris, Montague, and Cornwall in PE, and Digby, Yarmouth, Shelburne, and New Glasgow in NS, and Robert's Arm in NL.

Table 1: Summary of the findings: the found significant $R^2$ and/or $\rho$ between factors; sites[1] refer to sites with the highest 10% of FPC1 scores and the highest 10% of FPC2 scores (those in Fig. 3a); sites[2] refer to those with the highest 10% of FPC1 scores and the lowest 10% of FPC2 scores (those in Fig. 3b).

| FPC1 Score vs. Maximum FC Bacteria Level in | BC | QC, NB, PE, and NS | NL |
|---|---|---|---|
| | $R^2=0.61$<br>$p<2.2e-16$ | $R^2=0.68$<br>$p<2.2e-16$ | $R^2=0.74$<br>$p<2.2e-16$ |
| Approximated FC Bacteria Level at high-amplitude sites with strong seasonality in BC vs. | Meltwater-driven Water Flow at sites[1] | | Approximated Precipitation at sites[2] |
| | The Squamish River: $\rho \geq 0.74$<br>$p<2.2e-16$<br>The Fraser River: $\rho \geq 0.62$<br>$p<1.4e-06$ | | $\rho \geq 0.54$<br>$p<5.7e-05$ |

## 4. Discussion

This paper illustrates the main variation patterns in the fecal coliform bacteria levels on a one-year scale and their correlations with surrounding geographical factors at sites across coastal areas in Canada. Functional principal component analysis indicated that over 95% of the variance in the bacteria levels at sites across the 6 provinces could be decomposed into differences in the overall amplitude by FPC1 and the seasonality by FPC2, respectively. The former is correlated with regional human/animal population activities, while the latter with the precipitation around urbanized areas and with the discharge of meltwater-driven rivers elsewhere. These findings are the most obvious in British Columbia where the data are available throughout the year and similar in the other five provinces along the Atlantic coast: Quebec, New Brunswick, Prince Edward Island, Nova Scotia, and Newfoundland and Labrador, where the data in winter and early spring are not available. A summary of the findings is presented in Table. 2.

Regions with more human activities, such as population centers and areas with seasonal tourist activities, such as the west coast of Vancouver Island, tended to have relatively higher fecal coliform contamination. The relationship between the fecal contamination level and human population has been acknowledged in many studies. For example, a statistically significant positive correlation between fecal coliform levels and human population has been reported in the headwaters of the May River, South Carolina (Soueidan et al., 2021), and more fecal indicator bacteria are recorded at more central urban outlets, in the coastal zone of Gabon, Central Western Africa (Leboulanger et al., 2021). The insufficient treatment of human fecal matter appears to be a major cause for such phenomenon (Kataržytė et al., 2018; Durand et al., 2020). The biggest source of fecal pollution at Lake Pontchartrain, a tourist attraction, is also human associated in Louisiana (Xue et al., 2017). Moreover, the distribution of the contaminants across the coast is affected by coastal currents in Todos Santos Bay, Mexico (Tanahara et al., 2021), which could also be the case around Texada Island in the Strait of Georgia in British Columbia: according to the average surface circulation in the Strait of Georgia (Thompson, 1981), the amplitudes of fecal contamination at these sites might be brought up by some of the strongest surface circulation currents in the strait that carried the highly contaminated water from the surrounding cities.

Regarding the seasonality of the fecal contamination, in British Columbia, the staggering periods of high amounts of meltwater and precipitation allowed us to find regions with distinct fecal contamination seasonality correlated with meltwater or precipitation. Around urban areas, storm water or high precipitation can cause an overflow and worse water treatment than usual (Hall et al., 1998). In BC, precipitation in late fall and winter was usually very high, and certain locations suffered from inflow and infiltration where rainwater leaked into the sanitary sewer and could overwhelm the treatment system. Some moderately populated areas that coincided with higher overall amplitudes of fecal contamination used onsite wastewater systems for sewage disposal, which has been shown to relate to fecal contamination in the surrounding areas (Carroll et al., 2005). These systems could be sensitive to heavy rainfall, resulting in more contamination reaching marine waters (Humphrey et al., 2018). Meanwhile, runoff caused by high rainfall has been widely recognized as a strongly correlated factor to the fecal contamination in literature. For example, during wet seasons, the fecal coliform levels were significantly higher in Pacific Northwest estuaries (Zimmer-Faust et al., 2018), around the Jaranman Saryangdo area in Korea (Mok et al., 2016; Jung et al., 2017), and at two Costa Rican beaches (Laureano-Rosario et al., 2021). Runoff caused by precipitation might also result in flooding of the watershed and fecal matter on the surface being washed into the river (Boithias et al., 2020), such as the areas with upland agricultural activities along the east coast of Vancouver Island; the riverbank filtration has been proven important for reduction of the fecal pollution (Wang et al., 2021), so we also suspect that it could be the case that, with enough rainfall, the riverbank height is exceeded, and a pulse of high turbidity water may be washed into the marine waters, leading to high fecal coliform levels, as suspended sediments can play a significant role in the transportation of the bacteria (Chen and Liu, 2017). We infer that runoff elsewhere that was driven by meltwater-driven river discharge should have similar effects except for influences on sewage discharge systems. Meltwater-driven rivers like the Squamish River and the Fraser River in BC could carry substantial amount of fecal matter from upstream areas into marine waters in summer, potentially leading to peaks of bacteria levels at more distant locations. Fecal pollution entering the coastal marine environments in these approaches could also be influenced by the surface water circulation.

The seasonality pattern was not sufficiently clear along the Atlantic coast because the fecal coliform data in high precipitation seasons were not available. Regarding the precipitation patterns in the five provinces along Atlantic coast, high precipitation usually occurred in the spring and fall. Even though it was not as drastically dry as in BC, the precipitation in these provinces was relatively low in the summer as compared to the other seasons in the regions. Nevertheless, due to small numbers of fecal coliform measurements in winter and early spring there, we were unable to model the possible association between fecal bacteria levels and the precipitation. Thus, our result should not be interpreted as no association between precipitation and fecal bacteria level. For specific sites, there could be significant correlation between precipitation and fecal bacterial levels, especially at the sites with high FPC1 scores. Sites with significantly positive correlations between approximated levels of the two in each of the 5 provinces were grouped into 10 bins and plotted in Supplementary Fig. 11-15.

Additional analyses have been done on other factors in British Columbia. Fecal coliform bacteria levels have been found negatively correlated with salinity levels (Aslan-Yılmaz et al., 2004; Ortega et al., 2009; Soueidan et al. 2021), which agrees with our findings (data not shown). However, salinity could change as a result of rainfall (Bonilla et al. 2007), of which the situation was similar for temperature: it has a similar variation pattern as the discharge of meltwater-driven rivers.

In conclusion, the spatial and temporal fluctuations of marine pollution by fecal coliform were modelled using Functional Principal Component Analysis. The amplitude and seasonal variations of the fecal coliform pollutant are revealed in the functional principal components one and two, respectively; they are closely associated with fluctuation in the activities of human and warm-blooded animals. Precipitation and river discharges were the main vehicles to transport fecal material to the marine environment. This study characterizes the level of risk at particular sites and provide evidence for designing future monitoring or harvesting strategies. Closures and decontamination at specific sites after such risk evaluation and follow-up confirmation through additional sampling can be further implemented.

# Funding

This research was supported by grants DHGA-112-1 to YP and XZ from the Digital Health and Geospatial Analytics program of the National Research Council Canada, that reviewed the grants and provided funding for salaries, operational costs and services, but otherwise did not contribute intellectually to conceiving of the idea or interpreting the data.

# Acknowledgment

The authors thank Guillaume Durand, Colin Bellinger and Julio Valdes from NRC for their insights in the discussion as the project progressed, Yannick Losier for his assistance at the early stage of preliminary data analysis, and Zhaoze Liu for his contribution to building the interactive website.

# Figure Captions

Figure 1: Functional Principal Component Analysis (FPCA) on preprocessed BC fecal coliform data: (a) the mean function, $\mu$, was calculated as the average bacteria levels at all sites in the weeks; the variance from the mean function at the sites were then decomposed into Functional Principal Components (FPCs); (b) the top 2 FPCs together explained 95% of the variance, with corresponding proportions of variance explained in the brackets.

Figure 2: Characteristics of FPC1: (a) the maximum measurement on at each site against the corresponding percentile of FPC1 score, (b) the maximum measurement at a site with an FPC1 score in the highest (red) or the lowest (blue) 10% against the week of occurrence.

Figure 3: Characteristics of FPC2, river discharge, and precipitation along BC coast: approximated bacteria levels at sites with the highest 10% FPC1 scores and the highest 10% (a) and lowest 10% (b) of FPC2 scores (corresponding to the sites with red and blue dots of Fig. 2b; the mean are shown as the dashed black curves); average weekly flow of the Squamish River from 1999 to 2018 (c) and the Fraser River from 1999 to 2016 (e); (d) weekly precipitation at the sites presented in (b), with the mean values shown as the solid black curve; (f) average flows of small local rivers on the east coast of Vancouver Island, between Victoria and Courtenay.

Figure 4: Distribution of FPC1 scores in BC: the sites under analysis are marked as colored points on BC provincial map via "ggmap" package in R software; the ordered scores of the sites are grouped into 10 bins of equal size, and sequential colors from blue to red are assigned to sites in corresponding bins for efficient visualization of the relative magnitudes of FPC1 scores.

Figure 5: Distribution of FPC2 scores in BC: in the same way as for FPC1 scores, sites with relative magnitudes of FPC2 scores, indicated by the sequential colors, are marked along BC coast.

Figure 6: Correlation between precipitation and fecal coliform bacteria levels: distribution of sites with significant correlation between their approximated bacteria and precipitation levels; the sites were grouped into bins according to corresponding p-values (sites with p-values smaller than 1e-10 were all grouped into bin No. 10).

Figure 7: Characteristics of FPC1 in the provinces along Atlantic coast: the maximum measurement at a site against the corresponding percentile of FPC1 score in the 4 provinces combined (a) and NL alone (c); the maximum measurement at a site with an FPC1 score in the highest (red) or the lowest (blue) 10% against the week of occurrence in the 4 provinces combined (b) and NL alone (d).